\documentclass[lettersize,journal]{IEEEtran}
\usepackage{amsmath,amsfonts}
\usepackage{algorithmic}
\usepackage{array}
\usepackage[caption=false,font=footnotesize,labelfont=rm,textfont=rm]{subfig}
\usepackage{textcomp}
\usepackage{stfloats}
\usepackage{url}
\usepackage{verbatim}
\usepackage{graphicx}
\usepackage[colorlinks=true,linkcolor=blue]{hyperref}%
\usepackage{cite}
\def\BibTeX{{\rm B\kern-.05em{\sc i\kern-.025em b}\kern-.08em
    T\kern-.1667em\lower.7ex\hbox{E}\kern-.125emX}}
\usepackage{balance}
\begin{document}
\title{Generative Human Video Compression with Multi-granularity Temporal Trajectory Factorization}
\author{Shanzhi Yin, Bolin Chen~\IEEEmembership{Student Member,~IEEE}, Shiqi Wang~\IEEEmembership{Senior Member,~IEEE}, Yan Ye~\IEEEmembership{Senior Member,~IEEE} 
\thanks{Shanzhi Yin, Bolin Chen and Shiqi Wang are with the Department of Computer Science, City University of Hong Kong (E-mail: shanzhyin3-c@my.cityu.edu.hk, bolinchen3-c@my.cityu.edu.hk, shiqwang@cityu.edu.hk).} 
\thanks{ Yan Ye is with Damo Academy, Alibaba Group (E-mail: yan.ye@alibaba-inc.com).} 
 }


\markboth{SUBMITTED TO IEEE TRANSACTIONS ON CIRCUITS AND SYSTEMS FOR VIDEO TECHNOLOGY}%
{Generative Human Video Coding with Implicit Compact Motion Decomposition}

\maketitle

\begin{abstract}
In this paper, we propose a novel Multi-granularity Temporal Trajectory Factorization~(MTTF) framework for generative human video compression, which holds great potential for bandwidth-constrained human-centric video communication.
In particular, the proposed motion factorization strategy can facilitate to implicitly characterize the high-dimensional visual signal into compact motion vectors for representation compactness and further transform these vectors into a fine-grained field for motion expressibility. As such, the coded bit-stream can be entailed with enough visual motion information at the lowest representation cost.
Meanwhile, a resolution-expandable generative module is developed with enhanced background stability, such that the proposed framework can be optimized towards higher reconstruction robustness and more flexible resolution adaptation.
Experimental results show that proposed method outperforms latest generative models and the state-of-the-art video coding standard Versatile Video Coding (VVC) on both talking-face videos and moving-body videos in terms of both objective and subjective quality. The project page can be found at https://github.com/xyzysz/Extreme-Human-Video-Compression-with-MTTF.

\end{abstract}


\section{Introduction}
\IEEEPARstart{R}{ecnetly}, the blooming ``Short Video Era'' has witnessed the explosive growth of human-centric streaming media contents on many social networking applications. Therefore, ensuring efficient transmission and high-quality reconstruction of human videos is of paramount importance.
One of the solutions is to utilize generative human video coding~\cite{chen2024generativevisualcompressionreview}, which exploits strong statistical regularities of human contents and powerful inference capabilities of deep generative models to achieve superior Rate-Distortion (RD) performance compared to conventional hybrid codecs such as High Efficiency Video Coding (HEVC)~\cite{hevc} and Versatile Video Coding (VVC)~\cite{vvc}.
In particular, most of existing generative human video codecs are evolved from deep image animation methods~\cite{fomm,mraa,tpsm}, which could characterize the input high-dimensional visual signal into compact representations and employ the powerful deep generative model to achieve high-quality signal reconstruction/animation. 
For example, Deep Animation Codec~\cite{dac} utilizes 2D key-point representation for ultra-low bit-rate video conferencing. Similarly, 3D key-point is leveraged in talking-face video coding for free-view control~\cite{fv2v}, while feature matrices can represent facial temporal trajectory in a more compact manner~\cite{cfte}. 

However, the capabilities of existing generative human video coding schemes are limited by their feature representation and flow-warped generation designs.
On the one hand, these generative human video codecs mainly use explicit feature representation to characterize human faces, lacking expressibility and generalizability to handle more complicated scenarios such as human body movements. 
Meanwhile, such representations with actual physical manifestation could cause unnecessary compression redundancy. 
On the other hand, due to the fact that these schemes usually utilize flow-warped generation for the given reference signal, non-human parts of video contents could be mistakenly attached to moving human parts of the video, causing distortions on the edge of main object.
Furthermore, the flexibility of generative human video codecs is restricted by feature warping at a fixed feature size, making them unable to handle inputs of different resolutions.

In view of these existing limitations, this paper proposes an generative human video compression framework with multi-granularity temporal trajectory factorization~(MTTF). 
The proposed framework is crafted specifically to boost the capabilities of generative human video coding by enhancing both their generalizability and robustness. 
In particular, it explores a novel high-level temporal trajectory representation that can evolve complex motion modelling and texture details into multiple-granularity features. 
Moreover, such multi-granularity feature representations is not tied to any physical forms and can adapt well to diverse human video contents. 
Additionally, the proposed framework is capable of handling multiple resolutions via the dynamic generator and stabilize animated human through a parallel generation strategy. As such, both high-efficiency compression and high-quality reconstruction of human videos can be realized with better flexibility and scalability. The main contributions of this paper are summarized as follows,



\begin{itemize}


\item We propose an generative human video compression framework that enjoys advantages in representation flexibility, reconstruction robustness, scenario generalizability and resolution scalability. As such, the proposed framework can warrant the service of high-quality video communication with promising performance in versatile scenarios.

\item We design a multi-granularity feature factorization strategy to explore the internal correlations between compact motion vectors and fine-grained motion fields. In particular, this strategy can well guarantee the representation compactness for economical bandwidth and motion expressibility for high-quality signal reconstruction.

\item We develop a resolution-expandable generator that can dynamically adapt its network depth and width to inputs of different resolution. Meanwhile, it can stabilize  animated-human contents and improve reconstruction robustness by generating foreground and background in a parallel manner.

\item The experimental results show that our method can achieve state-of-the-art Rate-Distortion~(RD) performance compared to existing deep generative models and conventional codec on both talking-face videos and moving-body videos. Besides, our multi-resolution models can maintain superior performance under different input resolutions.
\end{itemize}

\section{Related Works}

\subsection{Hybrid Video Coding}
With the development of  video coding technologies for more than 40 years, a series of hybrid video codecs have been standardized to achieve remarkable compression capability, including Advanced Video Coding~(AVC)~\cite{avc}, HEVC~\cite{hevc},and VVC~\cite{vvc}. Recently, the Joint Video Experts Team (JVET) of ISO/IEC SC 29 and ITU-T SG16 has been actively involved in the next-generation video codec to exceed VVC by continuously optimizing coding tools towards a new Enhance Compression Model~(ECM) reference software~\cite{ecm}. In addition, efforts have also been made to discover the capability of Neural Network-based Video Coding (NNVC)~\cite{nnvc} and optimize traditional coding tools~\cite{deep-intra,deep-inloop} towards higher compression efficiency. In this paper, we utilize conventional hybrid codec VVC to compress the key frames of videos, which can  achieve high coding efficiency for key frames and provide high-quality texture reference for the generation of subsequent inter frames.

\subsection{End-to-End Coding}
Different from hybrid video coding where various coding tools are separately designed and optimized, end-to-end coding models are jointly trained in a data-driven manner. Ball\'e et al. proposed a series of pioneering works by realizing transform-quantization-coding pipeline with convolution neural networks and variational auto-encoder~\cite{factorized,variational,joint}. Further developments of end-to-end image coding include transform networks~\cite{bao22,tcm,devil}, entropy models~\cite{channelwise,elic,multirate,hu2020coarse}, light-weight structures~\cite{yin2022exploring,zhang2023elfic,yang2021slimmable}, semantic coding~\cite{Rethinking,conceptual} and coding for machine~\cite{vcm,ScalableFace}. Inspired by deep image coding, DVC~\cite{lu2019dvc} is one of the pioneers of end-to-end video coding, where all coding tools are realized by deep neural networks.
Following DVC, DCVC~\cite{dcvc} integrates conditional coding with feature domain context, DCVC-TCM~\cite{dcvc-tcm} utilizes temporal context mining, DCVC-HEM~\cite{dcvc-hem} introduces efficient spatial-temporal entropy mode, and DCVC-DC~\cite{dcvc-dc} further increases the context diversity in both temporal and spatial dimensions, 
Recently, DCVC-FM~\cite{dcvc-fm} expands the quality range and stabilizes long prediction chain with feature modulation, which can outperform ECM~\cite{ecm} under Low-Delay-Bidirectional~(LDB) setting.
Despite their superior compression capacity, these deep video coding methods still focus on low-level feature designs and cannot realize ultra-low bit-rate, while our proposed generative video coding method utilizes high-level compact feature representation and powerful generation model for extreme low bit-rate compression.

\subsection{Generative Video Coding with Deep Animation}
\subsubsection{Deep Image Animation}

Deep image animation techniques~\cite{fomm,monkey-net,mraa} can transfer the temporal motion to a reference image and utilize deep generative models to synthesize the high-quality video sequence.
The pioneer works for deep image animation are Monkey-Net~\cite{monkey-net} and FOMM~\cite{fomm}. They utilize self-supervised key-points and their local affine transformations to estimate the motion trajectory between objects. Afterwards, a series of works are proposed to improve the motion estimation precision and generation quality. In particular, MRAA~\cite{mraa} uses Principal Component Analysis~(PCA) decomposition of local affine transformations to animate articulated objects, Motion Transformer~\cite{motion-transformer} leverages transformer network to estimate affine parameters. DAM~\cite{dam} proposes structure-aware animation by regulating key-points as anchors and non-anchors, and enforcing correspondence between them. Other formats of motion parameterization are also explored such as second order motion model~\cite{som}, thin-plate spline transformation~\cite{tpsm}, continuous piece-wise-affine transformation~\cite{continuous}, and latent orthogonal motion vector~\cite{lia}. Moreover, other visual representations such as 3D mesh~\cite{liquid}, facial semantics~\cite{PuppeteerGAN} and depth map~\cite{dagan} are also leveraged to improve the animation performance.


Despite their strong generation ability, directly transferring deep image animation model to generative video codec has an obvious drawback. The compressibility could be compromised if the feature representation is not carefully designed for video coding. In~\cite{oquab2021low}, explicit features including landmarks, key-points and segmentation maps are implemented for low-bandwidth video chat compression. And it is observed that different feature representations lead to significantly different amount of bandwidth requirement. In this paper, we leverage deep image animation pipeline to construct our generative video coding framework, where feature representation and motion estimation both play important roles in achieving promising performance. Therefore, they should be carefully designed under the philosophy of video coding to remove redundancy and improve accuracy as much as possible.

\subsubsection{Generative Video Coding}
Inspired by deep image animation methods, generative video coding integrates conventional video codec and deep animation model to realize more efficient and intelligent coding paradigm. The pioneer works includes DAC~\cite{dac} that transfers FOMM~\cite{fomm} into generative codec with dynamic intra frame selection. Compact temporal trajectory representation with 4$\times$4 matrix is introduced in CFTE~\cite{cfte} and CTTR~\cite{cttr} for talking face video compression. Following these efforts, HDAC~\cite{hdac} further incorporates conventional codec as a base layer that is fused with generative prediction, while RDAC~\cite{rdac} incorporates predictive coding in generative coding framework. Besides, multi-reference~\cite{dr-cfte}, multi-view~\cite{multiview} and bi-directional prediction~\cite{bi-direction} schemes are also adopted to improve generation quality. To meet the requirements of more intelligent and practical application, FV2V~\cite{fv2v} allows free-view head control for video conferencing and IFVC~\cite{ifvc} provides more general interaction with the intrinsic visual representations. Recently, JVET has established a new ad-hoc group to establish testing conditions, develop software tools, and study compression performance and interoperability requirements for Generative Face Video Coding~(GFVC)~\cite{gfvc-review,gfvc,translator}, shedding light on the potential of further standardization of generative video coding techniques. Despite their rapid development, most of these methods are primarily focused on face videos, which limits their generalizability. In this paper, we extend generative video coding to more diverse contents with more intricate motion patterns, enhancing the versatility and effectiveness of this paradigm.


\section{Proposed Method}

\subsection{Framework Overview}

\begin{figure*}[t]
    \centering
    \includegraphics[width=18cm]{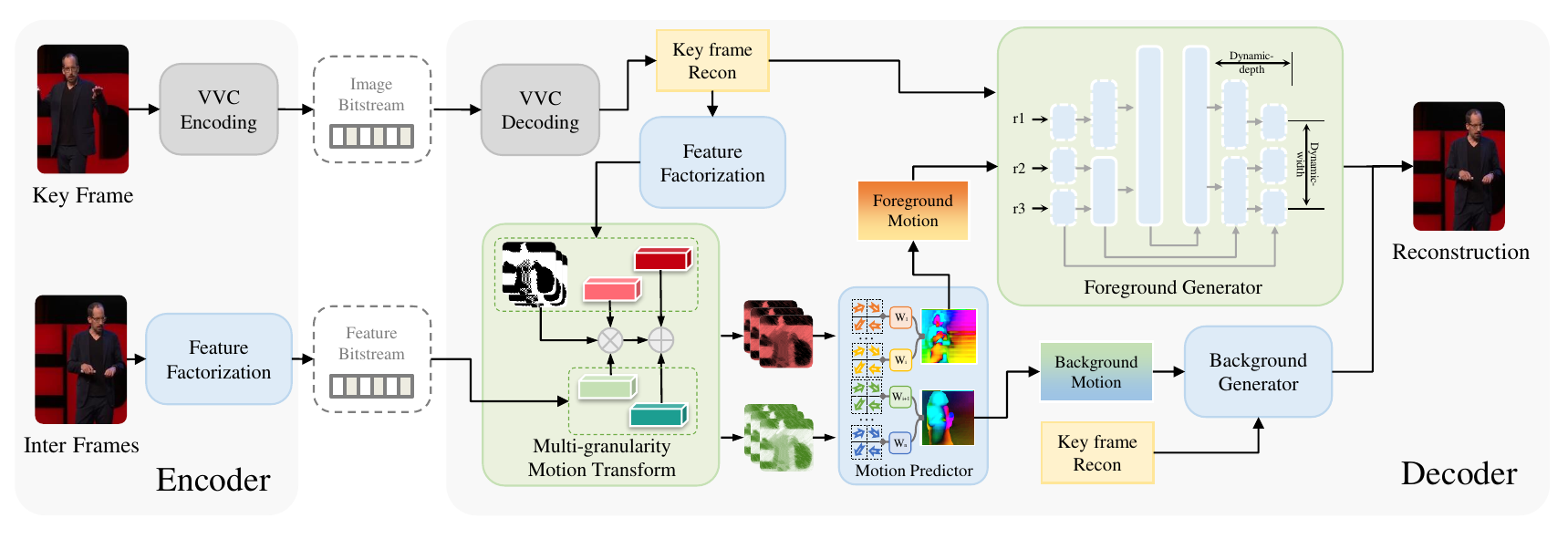}
    \caption{Overview of proposed generative human video coding framework.}
    \vspace{-0.26cm}
    \label{fig:1}
\end{figure*}

Our framework follows the general philosophy of generative face video coding~\cite{gfvc-review} and makes attempts to advance forward generative human video coding framework with richer video contents and better generation quality. As shown in Fig.~\ref{fig:1}, at the encoder side, the key frame (i.e., the first picture of the input sequence) is compressed by the conventional VVC codec and transmitted as an image bit-stream. Compact motion vectors are factorized from the subsequent inter frames and transmitted as feature bit-stream. To further reduce the feature redundancy between adjacent frames, we implement predictive coding following the practice in~\cite{cfte,cttr} and the predicted residuals are coded by Context-Adaptive Binary Arithmetic Coding~(CABAC).

At the decoder side, the key frame is first reconstructed by VVC codec, and then factorized to a spatial key latent and two compact motion vectors. For the inter frames, the compact motion vectors can be obtained by context-based entropy decoding and feature compensation from the feature bit-stream. Afterwards, these reconstructed compact motion vectors can be utilized to transform the spatial key latent, thus obtaining fine-grained motion fields. Specifically, each group of two motion vectors from key frame or inter frame are served as modulation weights and biases to perform spatial feature transform to the key latent. As such, the temporal trajectory information from two frames can be implicitly factorized into multi-granularity representations, i.e., compact motion vectors and fine-grained motion fields, by exploring their internal correlations with spatial feature transform. After that, fine-grained motion fields are fed into motion predictor to predict the sparse motion components and their weights. Then the sparse motion components are split and weighted-summed to form dense motions for foreground and background. Finally, the foreground and background are independently generated by resolution-expandable generators using the reconstructed key frame and corresponding motions. With input of different resolutions, the resolution-expandable generators can dynamically adjust their width and depth, such that the reconstructions can be compatible with different resolutions.

\subsection{Multi-granularity Temporal Trajectory Factorization}
\label{Multi-granularity Temporal Trajectory Factorization}
\begin{figure}[t]
    \centering
    \includegraphics[width=9cm]{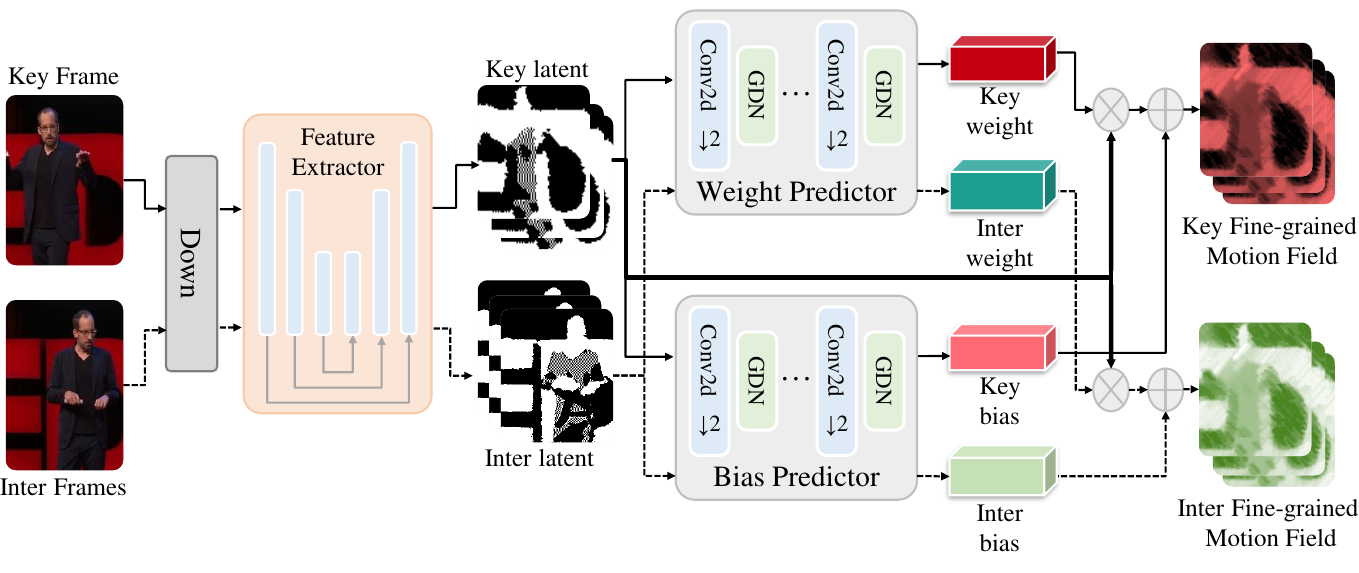}
    \caption{The detailed diagram of multi-granularity temporal trajectory factorization.}
    \label{fig:2}
    \vspace{-0.26cm}
\end{figure}

Feature representation is essential to generative coding, which should be concise enough for compact compression at the encoder side and informative enough for vivid generation at the decoder side.
For the existing deep image animation methods, they widely adopt explicit representations that are semantically related to video contents, such as 2D key-point~\cite{fomm,mraa,tpsm}, 3D key-point~\cite{fv2v} and segmentation map~\cite{PuppeteerGAN}.
These representations are not design for compression and may cause higher band-width costs~\cite{oquab2021low}. Recently, implicit feature representations that directly indicate motion information are proposed for generative compression/animation. In particular, CFTE~\cite{cfte} leverages a 4$\times$4 matrix to represent temporal trajectory evolution, while LIA~\cite{lia} extracts weight parameters for learned motion components. Herein, we propose a novel Multi-granularity Temporal Trajectory Factorization~(MTTF) scheme by considering both the compressibility and expressibility of trajectory representations and exploring the internal correlations between compact motion vectors and fine-grained motion fields.

We denote the reconstructed key frame and inter frame as $\hat{\textbf{I}}\in \mathbb{R}^{3\times H \times W}$ and $\textbf{P}\in \mathbb{R}^{3\times H \times W}$. They are first down-sampled by a ratio $s$ and fed into a feature extractor $E_{F}$ to obtain key latent and inter latent respectively,
\begin{equation}
    \textbf{L}_{\hat{I}} = E_{F}(D(\hat{\textbf{I}},s)),
\end{equation}
\begin{equation}
    \textbf{L}_{P} = E_{F}(D(\textbf{P},s)),
\end{equation}
where $D$ denotes down-sample operation and $\textbf{L}_{\hat{I}}$ and $ \textbf{L}_{P}$ are key latent and inter latent that share the dimension of ${N_{F}\times H/s \times W/s}$ and $N_{F}$ is the number of latents. Here, the key frame and inter frame share the same feature extractor. We implement a U-Net~\cite{unet} like structure, which contains down-sampling encoder, up-sampling decoder and short-cut concatenation from encoder to decoder. Then, each latent is fed into a weight predictor $E_{W}$ and a bias predictor $E_{B}$ to obtain compact motion vectors,
\begin{equation}
\textbf{w}_{\hat{I}} = E_{W}(\textbf{L}_{\hat{I}}),
\end{equation}
\begin{equation}
\textbf{b}_{\hat{I}} = E_{B}(\textbf{L}_{\hat{I}}),
\end{equation}\begin{equation}
\textbf{w}_{P} = E_{W}(\textbf{L}_{P}),
\end{equation}\begin{equation}
\textbf{b}_{P} = E_{B}(\textbf{L}_{P}),
\end{equation}
where $\textbf{w}_{\hat{I}}$, $\textbf{b}_{\hat{I}}$, $\textbf{w}_{P}$, $\textbf{b}_{P}$ are weight vectors and bias vectors from key frame and inter frame respectively and share the dimension of $N_{F} \times 1$. Here, we implement two independent predictors with the same structure, where down-sample layers and Generalized Divisive Normalization Layers~\cite{factorized} are cascaded to compress the latents to vectors.
Finally, we implement multi-granularity motion transform by modulating the key latent with weights and biases following the practice of spatial feature transform~\cite{stf},
\begin{equation}
    \textbf{F}_{\hat{I}} = \textbf{w}_{I} \cdot \textbf{L}_{\hat{I}} + \textbf{b}_{I},
\end{equation}
\begin{equation}
    \textbf{F}_{P} = \textbf{w}_{P} \cdot \textbf{L}_{\hat{I}} + \textbf{b}_{P},
\end{equation}
where $\textbf{F}_{\hat{I}}$ and $\textbf{F}_{P}$ are fine-grained motion fields for key frame and inter frame that share the dimension of $ {N_{F}\times H/s \times W/s}$, and $\cdot$ denotes channel-wise multiplication.

The detailed structure of multi-granularity temporal trajectory
factorization is illustrated in Fig.~\ref{fig:2}, where the input frames are factorized not only towards higher dimensions but also towards more diverse representations. The philosophy behind this mechanism is three-fold. First, we factorize the input signals into multiple channels of trajectory representations so that each dimension has the capability to implicitly describe different motion information. Due to the learnable, spatial-wise and input-adaptive feature design, MTTF enjoys better expressibility and flexibility than LIA~\cite{lia}, which only uses one set of motion vectors for all inputs.
Second, we use key frame latent as the basis of fine-grained motion fields without introducing additional coding bits. In comparison with CFTE~\cite{cfte} that only uses very compact 4$\times$4 matrix to describe motion changes, our employed key frame latent can provide more appearance information for motion representation.
Finally, only the inter-frame compact motion vectors, which serve as the transform coefficients for multi-granularity motion transform, need to be coded and transmitted. This ensures that the coded information remains compact enough to meet the requirements of ultra-low bit-rate coding.



\subsection{Coarse-to-fine Motion Estimation} \label{Motion Estimation}
After obtaining the fine-grained motion fields from reconstructed key frame and inter frames, the dense motion estimation process could be further executed in a coarse-to-fine manner. First, we estimate multiple motion components from given fine-grained motion fields using a flow predictor $FL$,
\begin{equation}
    \textbf{f} = FL(concat[\textbf{F}_{\hat{I}}, \textbf{F}_{P}]),
\end{equation}
where $\textbf{f}\in \mathbb{R}^{2N_{F} \times H/s \times W/s \times 2}$ denotes predicted coarse motion flow containing $2N_{F}$ components and $concat$ denotes concatenation operation. Here, we implement $FL$ as a U-Net like structure similar to $E_{F}$ and the predicted coarse motions are represented in the format of flow-field coordinate grids. Then, down-sampled key frame reconstruction is deformed by coarse motions,
\begin{equation}
    \hat{\textbf{I}}_{deformed} = Grid(D(\hat{\textbf{I}}),\textbf{f}),
\end{equation}
where $Grid$ denotes grid sample operation, and $\hat{\textbf{I}}_{deformed}$ denotes deformed key frame with the dimension of $ 3 \times 2N_{F} \times H/s \times W/s $. 
To combine the coarse motion components into finer dense motion, a weight predictor $W$ futher takes fine-grained motion fields and deformed key frames as inputs,
\begin{equation}
    \textbf{w}_{m} = W(\textbf{F}_{\hat{I}}, \textbf{F}_{P},\hat{\textbf{I}}_{deformed}),
\end{equation}
where $\textbf{w}_{m}$ denotes predicted weights with the dimension of $2N_{F} \times H/s \times W/s$. Here, we implement the weight predictor $W$ with a U-Net like structure. Then, to independently model the motion for foreground and background contents, the coarse motion and the corresponding weights are split into two parts, each containing $N_{fg}$ and $N_{bg}$ dimensions,
\begin{equation}
    \textbf{f}_{fg}, \textbf{f}_{bg} = split(\textbf{f}),
\end{equation}
\begin{equation}
    \textbf{w}_{fg}, \textbf{w}_{bg} = split(\textbf{w}).
\end{equation}

Finally, the motion weights are softmaxed with their own part and each motion part is weight-summed by corresponding weights,
 \begin{equation}
     \textbf{m}_{fg} = \sum_{i=1}^{N_{fg}} softmax(\textbf{m}_{fg})[i,:] \odot \textbf{f}_{fg}[i,:],
 \end{equation}
  \begin{equation}
     \textbf{m}_{bg} = \sum_{i=1}^{N_{bg}} softmax(\textbf{m}_{bg})[i,:] \odot \textbf{f}_{bg}[i,:],
 \end{equation}
where $\textbf{m}_{fg}$ and $\textbf{m}_{bg}$ denote the final dense motions of foreground and background as optical flow grid with the dimension of $H/s \times W/s \times 2$. $\odot$ denotes Hadamard product. Besides, we also predict an occlusion map $\textbf{occ}_{fg}$ for foreground generation from weight predictor,
\begin{equation}
    \textbf{occ}_{fg} = W(\textbf{F}_{\hat{I}}, \textbf{F}_{P},\hat{\textbf{I}}_{deformed}).
\end{equation}

\subsection{Foreground-and-background Parallel Generation}
\label{generation}
\begin{figure}[t]
    \centering
    \includegraphics[width=9cm]{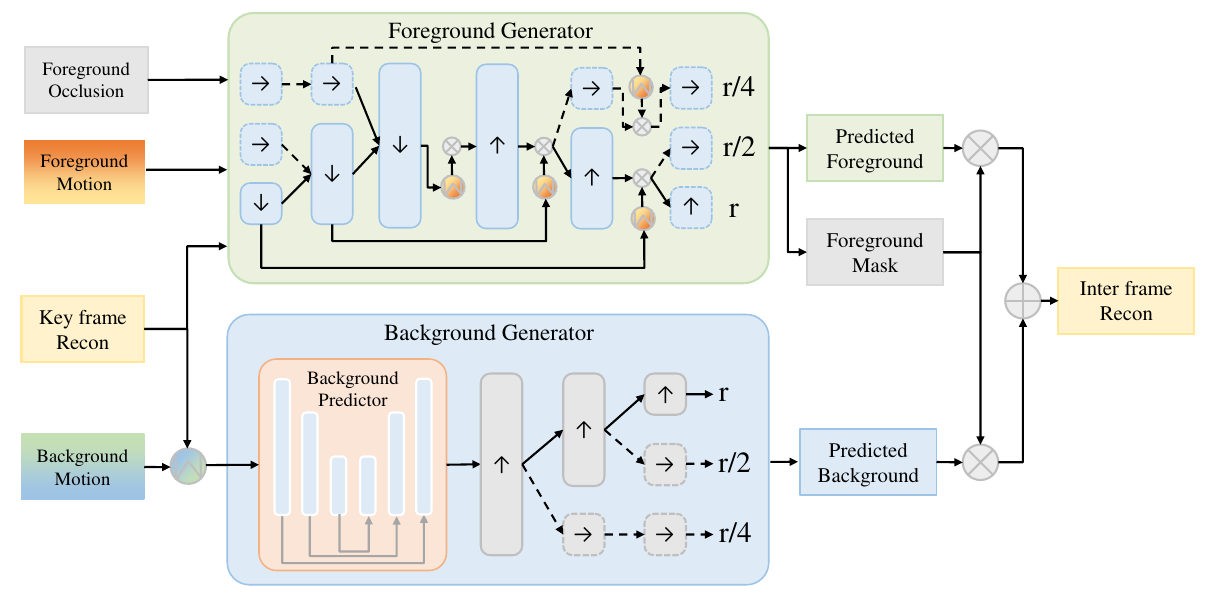}
    \caption{The detailed network structure of resolution-expandable generator. ``$\uparrow$'' denotes up-sample block, ``$\downarrow$'' denotes down-sample block and ``$\rightarrow$'' denotes blocks that maintain the feature size. ``w'' denotes warping with motion, ''$\times$'' denotes masking with occlusion. The network structure will be automatically initialized according to the depth and width setting. For example, the depth of network in this figure is set as 3. And if the width is 1, only modules with solid outline will be initialized. If the width is 3, all modules in the figure will be initialized.}
    \label{fig:3}
    \vspace{-0.26cm}
\end{figure}
In generative video coding, video contents often focus on common themes such as talking face or moving body. The movements of foreground object are larger than the background, which is often static or slightly shifting with camera. 
In previous deep image animation works, background motion is modeled independently using additional parameters~\cite{fomm,mraa,tpsm}, which can increase redundancy, while joint generation of background content with the foreground can introduce more distortions, potentially compromising both compressibility and reconstruction quality in generative coding.



To improve the stability of generation, we reverse the previous ``independent feature extraction, joint generation'' paradigm to the proposed ``joint feature extraction, independent generation'' paradigm.  In previous motion estimation, dense motions are already split into background motion and foreground motion. Then, key frame and background motion are utilized to generate background part in a ``warp-then-generate'' manner. Meanwhile, foreground motion and occlusion are utilized to generate foreground part and predict foreground mask in a ``warp-while-generate'' manner. Finally, we fuse the foreground generation and background generation to obtain final reconstruction. The detailed network structure is shown in Fig.~\ref{fig:3}.

\subsection{Resolution-expandable Generator}
To further improve the adaptivity and flexibility of generative video coding, we propose a resolution-expandable generator that can dynamically adjust its network width and depth to adapt to inputs of different resolutions. In deep image coding, input images are transformed to latent space for entropy coding~\cite{factorized,variational}. These latents are low-level features with image nuances that are compatible across different sizes. However, in generative video coding, features and motions are high-level information so that one model is normally trained and inferred on single resolution. Previously, CTTR~\cite{cttr} has explored multi-resolution generation using the traditional Bi-cubic frame interpolation algorithm, which is not flexible and could cause blurring artifacts. In the paper, we use more dynamic network structure to achieve higher multi-resolution scalability. 

We adopt resolution-expandable generator for both foreground generation and background generation. Without loss of generality, we denote the largest possible input resolution as $r$ and number of supported resolutions as $N_{s}$ with a multiplier of 2 between adjacent resolutions,
\begin{equation}
\label{ri}
    r_{i} \in  \{\frac{r}{2^{i}}, i=0,1..N_{s}-1\}.
\end{equation}
As described in section~\ref{Multi-granularity Temporal Trajectory Factorization}, there is a down-sample factor $s$ between motions $\textbf{m}_{fg}$, $\textbf{m}_{bg}$ and input images. During generation, the number of encoder or decoder blocks (depth) in generator is $N_{B} = \log_{2}s$ to match the size of motions. To handle all resolutions,  the width of generation will be $N_{s}$ and should not be larger than $N_{B}$. In Fig.~\ref{fig:3}, we give an example where $N_{s} = N_{B} = 3$.

Specifically, for background generation, we first warp the down-sampled the key frame reconstruction with background motion, and feed the output into a background predictor $BG$,
\begin{equation}
    \textbf{F}_{BG} = BG(D(\hat{\textbf{I}},s) \star \textbf{m}_{bg}),
\end{equation}
where $\star$ denotes warping operation and $\textbf{F}_{BG}$ denotes output background feature. Here, we design $BG$ as U-Net like structure. Then, the feature is processed by cascaded decoder blocks. According to the desired output resolution $r_{i}$, there should be $n_{u} = \log_{2}s\cdot \frac{r_{i}}{r}$ up-sample blocks in all $N_{B}$ blocks,
\begin{equation}
    \hat{\textbf{P}}_{bg} =\sigma ( (\textbf{g}_{N_{B}-n_{u}} .. \textbf{g}_{2} \circ \ \textbf{g}_{1} \circ \textbf{u}_{n_{u}} \circ ..\textbf{u}_{2}\circ \textbf{u}_{1})(\textbf{F}_{BG} ))
\end{equation}
where $\textbf{u}$ denotes up-sample block,  $\textbf{g}$ denotes normal decoder block that maintain the feature size, $\sigma$ denotes sigmoid activation and $ \hat{\textbf{P}}_{bg}$ denotes generated background for inter frame reconstruction.

For foreground generation, we first down-sample the key frame reconstruction to feature with the same size of foreground motion in the encoder part, and the feature is warpped by the foreground motion,
\begin{equation}
    {\textbf{F}}_{fg}^{0} =  ((\textbf{d}_{n_{u}} \circ ..\textbf{d}_{2}\circ \textbf{d}_{1}  \circ \textbf{g}_{N_{B}-n_{u}} .. \textbf{g}_{2} \circ \ \textbf{g}_{1})(\hat{\textbf{I}} )) \star \textbf{m}_{fg},
\end{equation}
where $\textbf{d}_{i}$ denotes down-sample block. Then, after every decoder block $\textbf{b}_{i}$, the feature is weighted-summed with warpped feature $\textbf{F}_{fg}^{-i}$ from the corresponding block of encoder part,
\begin{equation}
    \textbf{F}_{fg}^{i} = \textbf{b}_{i}(\textbf{F}_{fg}^{i-1}) \cdot (1-\textbf{occ}_{fg}) + (\textbf{F}_{fg}^{-i} \star \textbf{m}_{fg}) \cdot \textbf{occ}_{fg},
\end{equation}
where $\textbf{b}_{i}$ denotes block that would be $\textbf{u}$ if $i<n_{u}$ and otherwise would be $\textbf{g}$.
To save computation and improve efficiency of our resolution-expandable generator, all features with same input size shares a down-sample block during encoding, and all features with output size share an up-sample block during decoding. Inputs with different resolution will go through different routes in resolution-expandable generator as illustrated in Fig.~\ref{fig:3}.
Finally, we predict foreground reconstruction $ \hat{\textbf{P}}_{fg}$ and foreground mask $ \textbf{M}_{fg}$ from the last decoder feature,
\begin{equation}
    \hat{\textbf{P}}_{fg}, \textbf{M}_{fg} =\sigma ( split(\textbf{F}_{fg}^{N_{B}})).
\end{equation}
Finally, the inter frame reconstruction would be fused from foreground generation and background generation with the predicted mask,
\begin{equation}
    \hat{\textbf{P}} = \textbf{M}_{fg} \cdot \textbf{P}_{fg} +  (1- \textbf{M}_{fg})  \cdot \textbf{P}_{bg}.
\end{equation}

\subsection{Model Optimization}
We optimize the proposed method following the common practice in deep image animation and generative video coding.
\subsubsection{Perceptual Loss} 
To improve the generation quality, the perceptual-level reconstruction can be regularized by comparing features extracted by VGG-19 network~\cite{fomm}. Here, we use multi-scale reconstruction similar to~\cite{mraa,fomm},
\begin{equation}
    \mathcal{L}_{per} = \sum_{j=1}^{4} \sum_{i=1}^{5} \frac{||VGG_{i}(D(\hat{\textbf{P}},\frac{1}{2^{j}}))-VGG_{i}(D (\textbf{P},\frac{1}{2^{j}}))||}{C_{i} \cdot H_{i} \cdot W_{i}}, 
\end{equation}
where $VGG_{i}$ denotes feature from $i$th layer from VGG network with dimension of $C_{i} \times H_{i} \times W_{i}$ and $D$ denotes down-sample operation.
\subsubsection{L1 Loss} To further regulate pixel-level reconstruction, we also employ L1 loss on generated key frame,
\begin{equation}
    \mathcal{L}_{L1} =  \frac{||\hat{\textbf{P}}-\textbf{P}||_{1} }{C \cdot H \cdot W}.
\end{equation}
\begin{figure}[t]
    \centering
    \subfloat[Moving body test set.]
{\includegraphics[width=8.5cm]{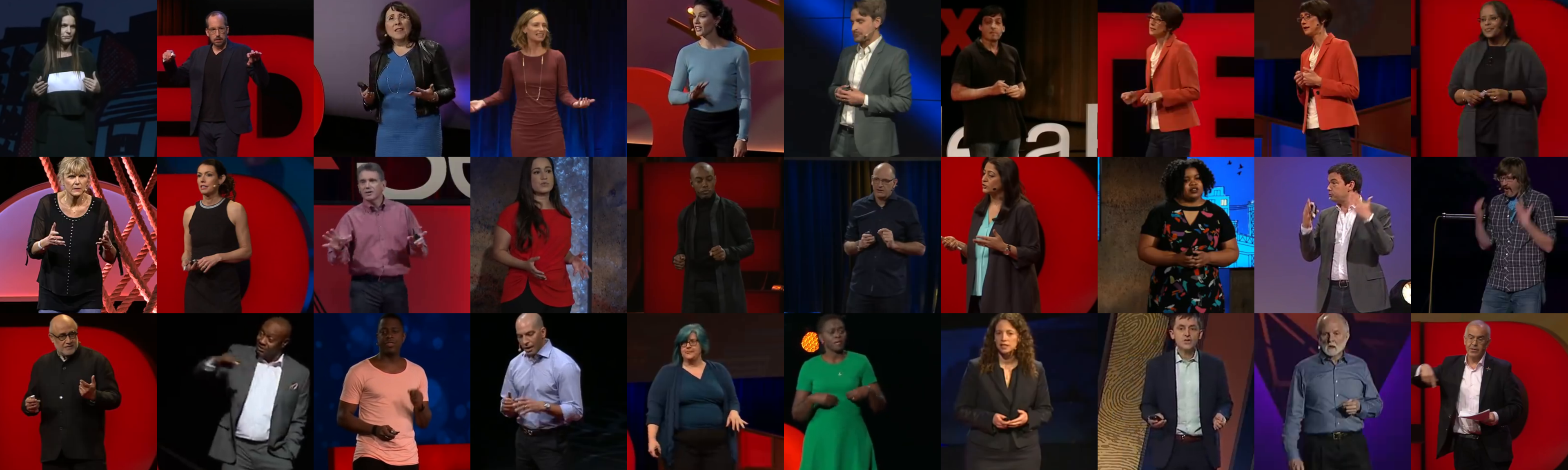}}
\\
\subfloat[Talking face test set.]{\includegraphics[width=8.5cm]{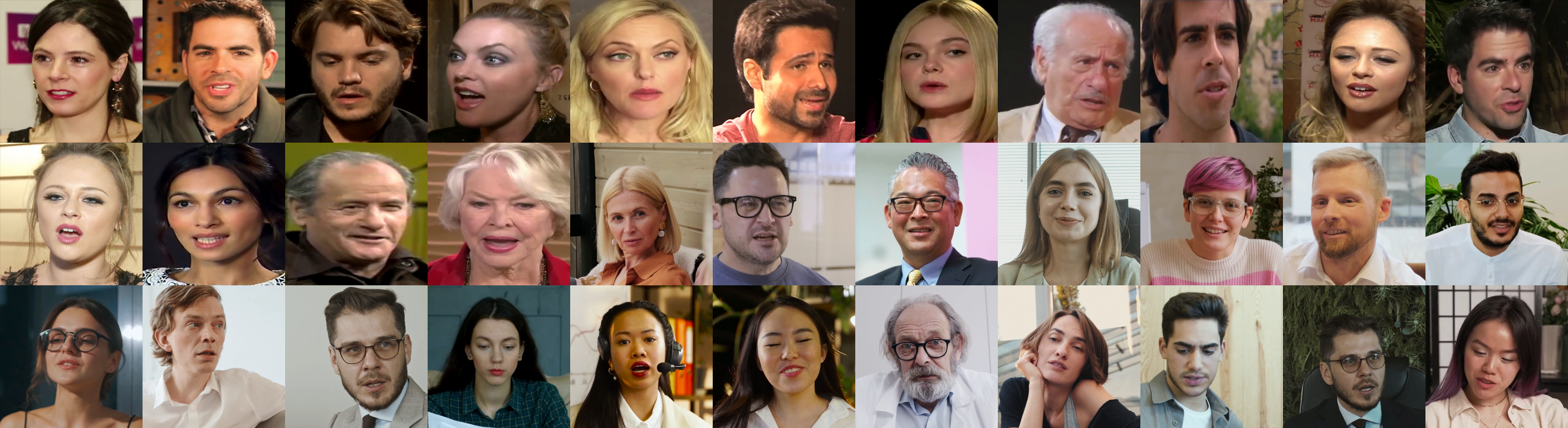}}

    \caption{Overview of test set.}
    \label{fig:45}
    \vspace{-0.26cm}
\end{figure}
\subsubsection{Background Loss} 
To ensure the independent generation of background and foreground, we use off-the-shelf image matting model~\cite{bg} to provide ground truth mask for generated mask $\textbf{M}_{fg}$.
\begin{equation}
    \mathcal{L}_{bg} =  \frac{||\textbf{M}_{fg}- \phi (\textbf{P})||_{1} }{C \cdot H \cdot W},
\end{equation}
where $\phi$ denotes the image matting model.

To sum up, the total objective for model optimization is 
\begin{equation}
   \mathcal{L} =  \lambda_{per} \cdot \mathcal{L}_{per} +\lambda_{L1} \cdot \mathcal{L}_{L1}+ \lambda_{bg} \cdot \mathcal{L}_{bg},
\end{equation}
where $\lambda_{per}$, $\lambda_{L1}$ and $\lambda_{bg}$ are weights for perceptual loss, L1 loss and background loss respectively and we empirically set them equally as 10.
\subsubsection{Multi-resolution Training}
To optimize our proposed resolution-expandable generator, we randomly select input resolution $r_{i}$ according to eq.~\ref{ri} and calculate losses across all resolutions,
\begin{equation}
     \mathcal{L}_{multiRes} = \sum_{i=1}^{N_{s}}  \mathcal{L}(\textbf{P}_{r_{i}},\hat{\textbf{P}}_{r_{i}})
\end{equation}
where $\textbf{P}_{r_{i}}$ denotes input frame of $r_{i}$ resolution.

\section{Experimental Results}
To verify the capability and generalizability of our proposed framework, our experiments are carried out for two different scenarios, i.e., moving human body videos and talking face videos.
\subsection{Experimental Settings}
\subsubsection{Datasets}
\begin{figure*}[t]
\centering
\subfloat[Rate-DISTS]
{\includegraphics[width=6cm]{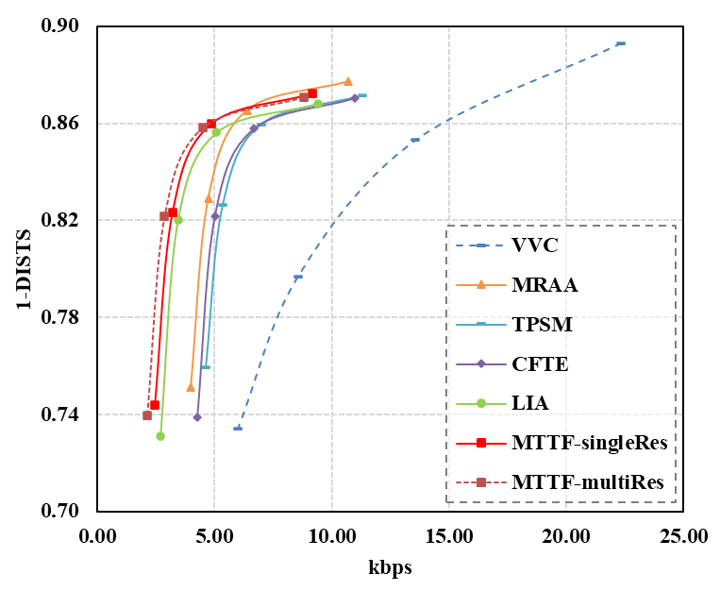}}
\subfloat[Rate-LPIPS]{\includegraphics[width=6cm]{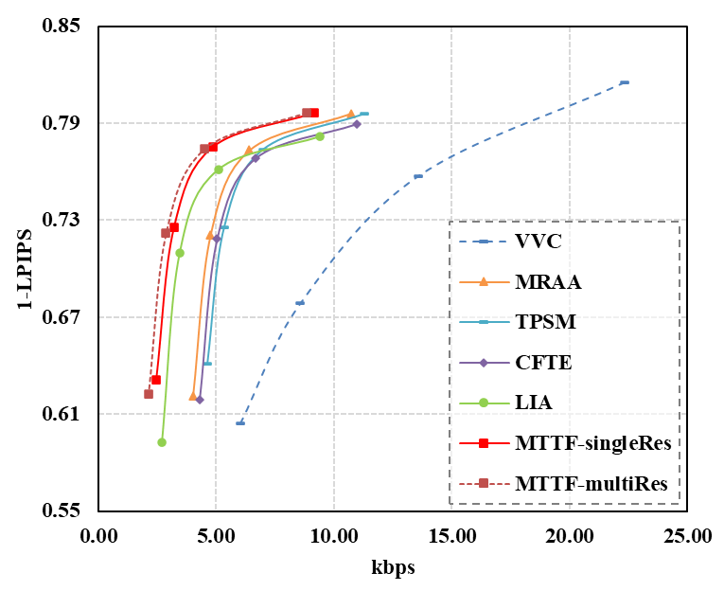}}
\subfloat[Rate-FVD]{\includegraphics[width=6cm]{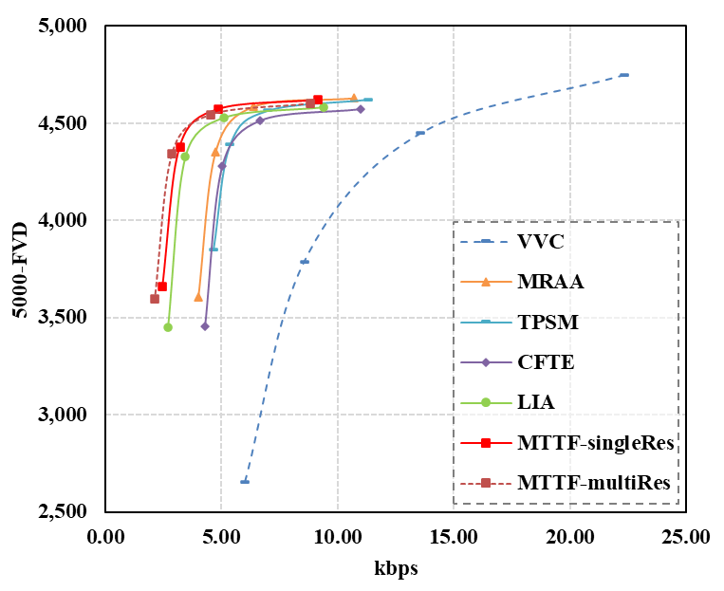}}
 \caption{Rate-distortion performance comparisons with VVC~\cite{vvc}, FOMM~\cite{fomm}, MRAA~\cite{mraa}, TPSM~\cite{tpsm}, CFTE~\cite{cfte} in terms of DISTS, LPIPS and FVD for moving-body test set.  }
 \vspace{-0.26cm}
 \label{fig:6}
\end{figure*}

\begin{figure*}[t]
\centering
\subfloat[Rate-DISTS]
{\includegraphics[width=6cm]{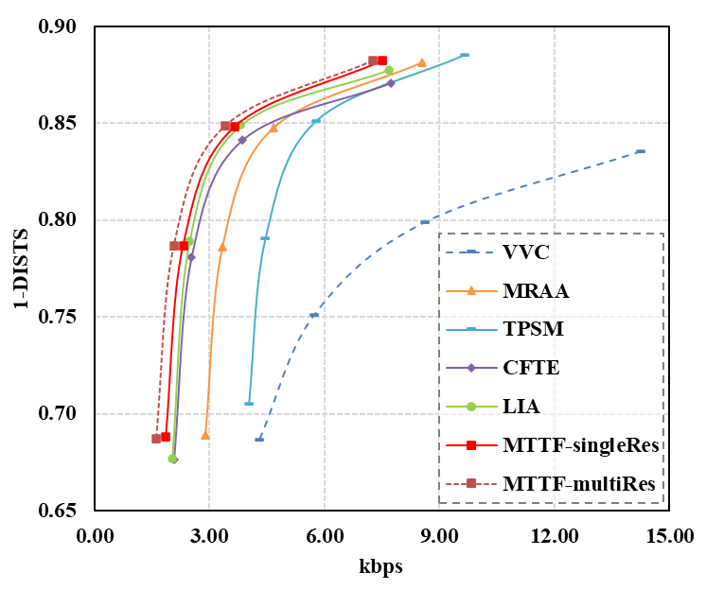}}
\subfloat[Rate-LPIPS]{\includegraphics[width=6cm]{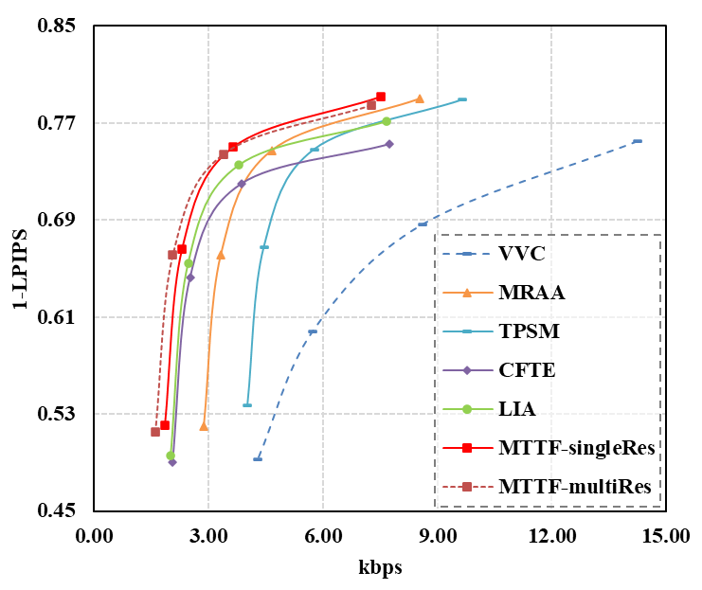}}
\subfloat[Rate-FVD]{\includegraphics[width=6cm]{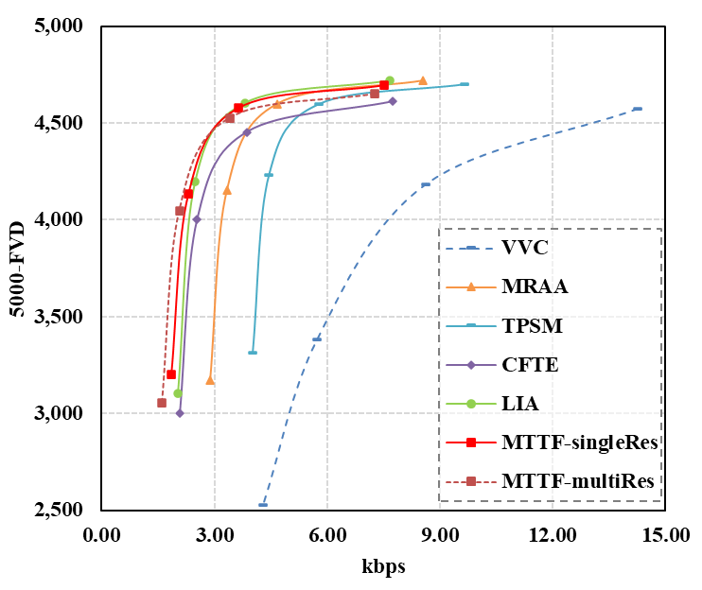}}
  \caption{Rate-distortion performance comparisons with VVC~\cite{vvc}, FOMM~\cite{fomm}, MRAA~\cite{mraa}, TPSM~\cite{tpsm}, CFTE~\cite{cfte} in terms of DISTS, LPIPS and FVD for talking-facetest set.  }
  \vspace{-0.26cm}
 \label{fig:7}
\end{figure*}

For moving human body scenario, we verify our method on TEDTalk dataset, which was first used in~\cite{mraa} for articulated objects animation. This dataset mainly contains 1132 training videos and 128 testing videos with the resolution of 384$\times$384. We train the proposed framework on TEDTalk training set. In addition, we select 30 videos with different identities from TEDTalk test set for evaluation, where each video contains 150 frames, as shown in Fig.~\ref{fig:45} (a). 

For talking face scenario, we train our method on VoxCeleb training dataset~\cite{vox}, mainly containing 18,641 training videos at the resolution of 256$\times$256. For evaluation, we follow the common test condition of GFVC~\cite{gfvc-ctc} and use the corresponding 33 testing sequences as shown in Fig.~\ref{fig:45} (b). Among them, 15 sequences have 250 frames with head-centric contents, and 18 sequences have 125 frames with head-and-shoulder contents.

\subsubsection{Implementation Details}

We implement our proposed generative compression algorithm with the Pytorch framework and use NVIDIA TESLA A100 GPUs for model training. In particular, these models are trained with 100 epochs via the Adam optimizer with $\beta_{1} = 0.5$, $\beta_{2}=0.999$. 
Besides, we set the initial learning rate as $0.0002$ and use multi-step learning rate scheduler with $\gamma=0.1$ and  $milestone=[60,90]$.
As for network parameter settings, the down-scale factor $s$ is set at 0.25. 
The number of feature dimension $N_{F}$ is set at 20, thus the number of motion components will be 40. According to empirical experiments,  the number of background $N_{bg}$ and foreground motion components   $N_{fg}$ are set at 5 and 35, respectively. 

Regarding the multi-resolution training, we choose the largest resolution $r=768$ for the TEDTalk dataset and $r=512$ for the VoxCeleb dataset. In addition, the supported resolution $N_{s}$ is further set at 3, i.e, our proposed multi-resolution model can support 768px/384px/192px resolutions for moving-body video compression and 512px/256px/128px resolutions for talking-face video compression. It should be noted that to better adjust different training data resolutions, the Lanczos operation is adopted for the frame interpolation in original training data.
Besides, we also train single-resolution models with fixed resolution of 384$\times$384 and 256$\times$256 for TEDTalk and VoxCeleb, respectively.


\subsubsection{Quality Evaluation Metrics}

For objective quality measurement, we follow the common practice in deep image animation~\cite{fomm,mraa,tpsm} and generative video coding~\cite{cfte,gfvc-ctc,cttr}. In particular, we choose two metrics for perceptual-level image quality assessment, i.e., Learned Perceptual Image Patch Similarity~(LPIPS)~\cite{lpips} and Deep Image Structure and Texture Similarity~(DISTS)~\cite{dists}. These two measures quantify the mean square error and structural similarity on feature maps extracted by VGG network, which may be appropriate for the evaluation of GAN-based compression. Additionally, we also choose Frechet Video Distance~(FVD)~\cite{unterthiner2019fvd} to evaluate the temporal consistency by capturing the temporal dynamics and comparing the feature distribution between the original/reconstructed videos. Bjøntegaard-delta-rate~(BD-rate)~\cite{bdrate} and   rate-distortion~(RD) curve are adopted to quantify the overall compression performances between the proposed codec and other compared anchors. For LPIPS, DISTS and FVD, the lower their values are, the better their perceived quality are. To display the RD curves in an increasing manner and calculate BD-rate saving, we use ``1-DISTS'', ``1-LPIPS'' and ``5000-FVD'' as the y-axis of the graphs.

\subsection{Compared Algorithms}

To verify the effectiveness of the proposed method, we select 1 state-of-the-art conventional video codec VVC~\cite{vvc} and 4 latest generative video codecs MRAA~\cite{mraa}, TPSM~\cite{tpsm}, CFTE~\cite{cfte}, LIA~\cite{lia} for comparisons. In the following, we discuss the implementation details. 

\subsubsection{Conventional VVC Codec} It is the latest hybrid video coding standard, which significantly improves the rate-distortion performance compared with its predecessors. We adopt the Low-Delay-Bidirectional (LDB) configuration in VTM 22.2 reference software for VVC, where the quantization parameters (QP) are set to 37, 42, 47 and 52.
\subsubsection{Generative Video Codecs}They are mainly rooted in deep animation models and are further migrated into generative video compression tasks, i.e., their corresponding feature extraction module and motion estimation with frame generation modules are regarded as the encoder and decoder. We choose four different generative models with three different feature representations, i.e., MRAA~\cite{mraa} and TPSM~\cite{tpsm} with key-points, CFTE~\cite{cfte} with compact matrices and LIA~\cite{lia} with weighting coefficients. In addition, other configurations like the key frame image and inter frame feature compression strictly follow our proposed pipeline. In particular, the key frame is compressed via the intra mode of VTM 22.2 software with the QPs of 22, 32, 42, 52, and inter frame features are compressed via a context-adaptive arithmetic coder.  The detailed implementations are described as follows,

\subsection{Performance Comparisons}
\subsubsection{Objective Performance}
\begin{figure*}[t]
\centering
\subfloat[Moving-body]
{\includegraphics[width=14cm]{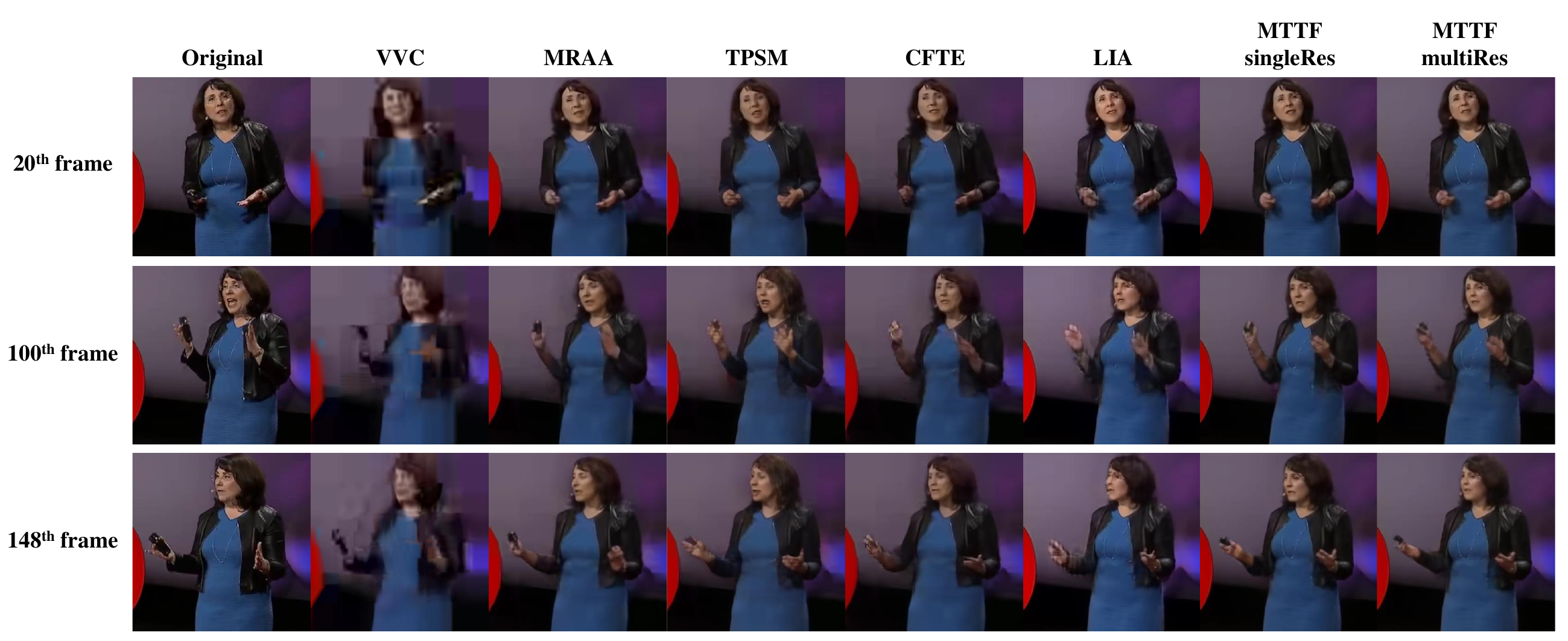}}
\\
 \vspace{-0.26cm}
\subfloat[Talking-face]{\includegraphics[width=14cm]{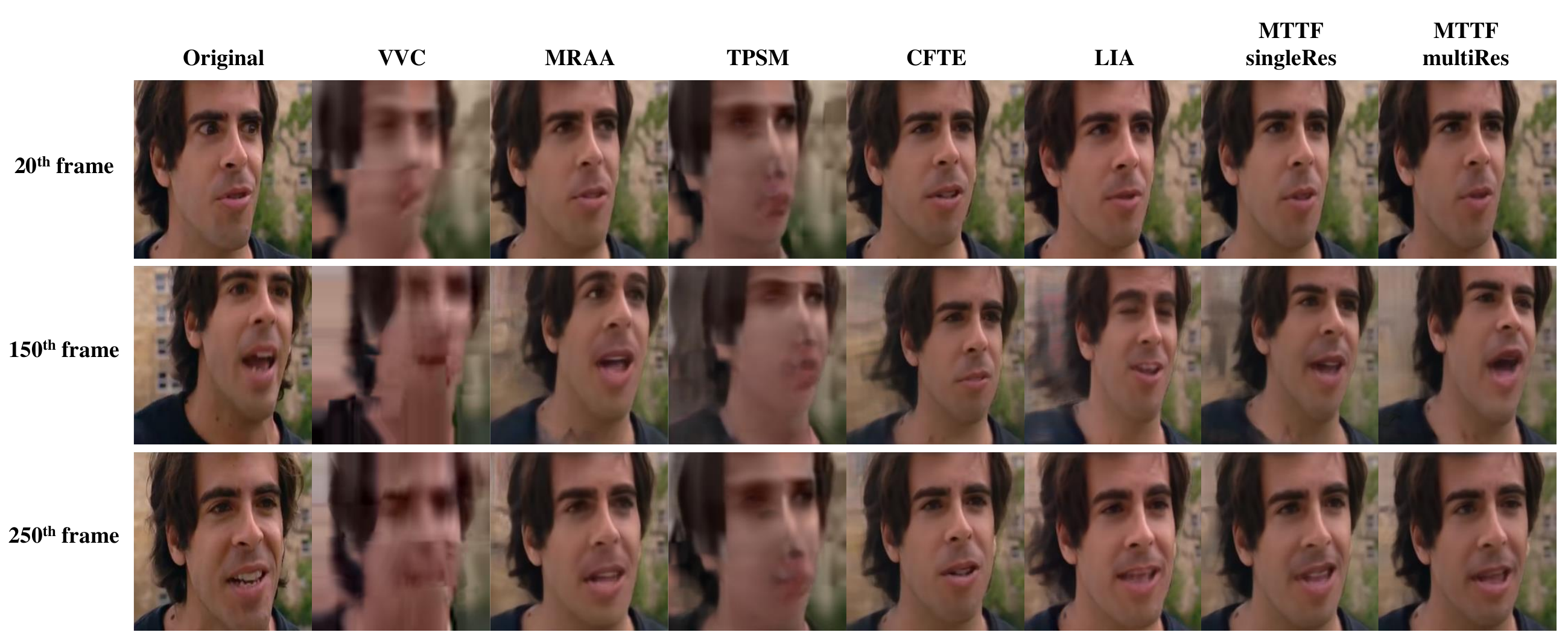}}
 \caption{Visual quality comparisons on moving-body and talking-face test sets among VVC~\cite{vvc}, FOMM~\cite{fomm}, MRAA~\cite{mraa}, TPSM~\cite{tpsm}, CFTE~\cite{cfte} and MTTF (Ours) at the similar bit rate of 5kbps.}
 \label{fig:11}
 \vspace{-0.26cm}
\end{figure*}

\begin{table}[t]
\caption{RD performance comparisons on moving-body test set (384$\times$384 resolution) in terms of average BD-rate savings over the VVC anchor~\cite{vvc}.}
    \label{tab:1}
    \centering
\renewcommand\arraystretch{1.1}
\begin{tabular}{cccc}
\hline
Algorithm      & Rate-DISTS & Rate-LPIPS & Rate-FVD \\ \hline
MRAA~\cite{mraa}         & -50.92\%            & -51.02\%            & -55.66\%          \\
TPSM~\cite{tpsm}           & -44.51\%            & -47.22\%            & -52.08\%          \\
CFTE~\cite{cfte}           & -44.72\%            & -47.10\%            & -49.68\%          \\
LIA~\cite{lia}            & -61.99\%            & -61.77\%            & -67.33\%          \\
\textbf{MTTF-singleRes} & \textbf{-65.96}\%            & \textbf{-68.16}\%            & \textbf{-71.27}\%          \\
\textbf{MTTF-multiRes}  & \textbf{-69.35}\%            & \textbf{-70.95}\%            & \textbf{-73.76}\%          \\ \hline
\end{tabular}
\vspace{-0.2cm}
\end{table}

\begin{table}[t]
\caption{RD performance comparisons on talking-face test set (256$\times$256 resolution)  in terms of average BD-rate savings over the VVC anchor~\cite{vvc}.}
    \label{tab:2}
    \centering
\renewcommand\arraystretch{1.20}
\begin{tabular}{cccc}
\hline
Algorithm               & Rate-DISTS        & Rate-LPIPS        & Rate-FVD          \\ \hline
MRAA~\cite{mraa}                    & -48.75\%          & -50.61\%          & -55.51\%          \\
TPSM~\cite{tpsm}                    & -35.33\%          & -36.27\%          & -42.17\%          \\
CFTE~\cite{cfte}                    & -59.93\%          & -58.49\%          & -62.30\%          \\
LIA~\cite{lia}                     & -62.21\%          & -61.36\%          & -67.01\%          \\
\textbf{MTTF-singleRes} & \textbf{-64.24\%} & \textbf{-65.82\%} & \textbf{-69.14\%} \\
\textbf{MTTF-multiRes}  & \textbf{-67.98\%}       & \textbf{-68.72\%}       & \textbf{-70.83\%}       \\ \hline
\end{tabular}
\vspace{-0.2cm}
\end{table}

\begin{table}[t]
\caption{Multi-resolution compression performance comparisons on moving-body test set in terms of average BD-rate saving against VVC~\cite{vvc}.}
    \label{tab:3}
    \centering
    \renewcommand\arraystretch{1.20}
\begin{tabular}{cccc}
\hline
Resolution & Rate-DISTS & Rate-LPIPS & Rate-FVD \\ \hline
192 $\times$ 192       &    -66.47\%        & -66.08\%            &   -73.98\%       \\
384  $\times$  384       & -69.35\%   & -70.95\%   & -73.76\% \\
768   $\times$  768      &  -64.76\%          &  -69.58\%          &   -70.08\%       \\ \hline
\end{tabular}
\vspace{-0.2cm}
\end{table}

\begin{table}[t]
\caption{Multi-resolution compression performance comparisons on talking-face test set dataset in terms of average BD-rate saving against VVC~\cite{vvc}.}
    \label{tab:4}
    \centering
    \renewcommand\arraystretch{1.20}
\begin{tabular}{cccc}
\hline
Resolution & Rate-DISTS & Rate-LPIPS & Rate-FVD \\ \hline
128  $\times$  128       & -63.57\%           & -63.15\%           &     -65.11\%     \\
256  $\times$ 256       &   -67.98\%           &  -68.72\%        & -70.83\%        \\
512   $\times$ 512      &   -69.32\%         &      -60.79\%      &   -72.42\%       \\ \hline
\end{tabular}
\vspace{-0.2cm}
\end{table}
Fig.~\ref{fig:6} shows the RD performance of our proposed MTTF method and different codecs in terms of Rate-DISTS, Rate-LPIPS and Rate-FVD on moving-body scenario. It can be seen that our proposed method can outperform all compared methods on three perceptual-level measures. Moreover, our multi-resolution model (MTTF-multiRes) even performs slightly better than our single-resolution model (MTTF-singleRes). The specific BD-Rate saving against VVC is shown in Table~\ref{tab:1}, illustrating that our proposed method can achieve promising BD-Rate saving with more than 70\% on Rate-LPIPS and Rate-FVD at the resolution of 384$\times$384. As illustrated in Fig.~\ref{fig:7}, our proposed models (i.e., MTTF-multiRes and MTTF-singleRes) also can achieve superior RD performance in comparisons with other compression algorithms on talking-face scenario. In addition, Table~\ref{tab:2} shows that our proposed MTTF-multiRes algorithm can achieve 67.98\% average bit-rate savings in terms of Rate-DISTS, 68.72\% bit-rate savings in terms of Rate-LPIPS and 70.83\% bit-rate savings in terms of Rate-FVD at the resolution of 256$\times$256.


Overall, it can be seen that our proposed MTTF method can show superior performance on both moving-body and talking-head scenarios, greatly illustrating both the generalizability and robustness of our framework. Thereafter, LIA~\cite{lia} achieves the second-best performance with similar bit-rate but slightly lower quality than ours, showing that our input-adaptive fine-grained motion fields have better flexibility and adaptivity.
On the contrary, MRAA~\cite{mraa} and TPSM~\cite{tpsm} show higher bit-rate cost because of their explicit key-point-based feature representation. For CFTE~\cite{cfte}, it achieves better performance on talking-face scenario than more complicated moving-body scenario, showing that its expressibility may be limited by its pure compact-feature-based design, while our proposed MTTF method utilizes both compact motion vectors and fine-graind motion fields for trajectory representations.

\subsubsection{Subjective Performance}
\begin{table*}[t]
\caption{USER PREFERENCE IN PAIRWISE COMPARISON IN TERMS OF SIMILAR CODING BITS CONSUMPTION.}
    \label{tab:5}
    \centering
\renewcommand\arraystretch{1.20}
\begin{tabular}{ccccccc}
\hline
Algorithm   Comparisons & Bitrate~(kbps) & DISTS~($\downarrow$) & LPIPS~($\downarrow$) & FVD~($\downarrow$) &   PSNR~($\uparrow$) & User Preference \\ \hline
VVC~\cite{vvc} / Ours               &   4.85 / 4.21      &     0.32 / 0.14  &   0.49 / 0.23     &  2345.29 / 390.17   &23.63 / 24.74 & 0.00\% / \textbf{100.00}\%              \\
MRAA~\cite{mraa} / Ours              &  4.20 / 4.21       &    0.20 / 0.14   &   0.32 / 0.23    &   784.82 / 390.17  &24.15 / 24.74 & 7.50\% / \textbf{92.50}\%              \\
TPSM~\cite{tpsm} / Ours              &   4.74 / 4.21      &     0.24 / 0.14  &   0.38 / 0.23     &  1079.12 / 390.17   &22.70 / 24.74 &  5.00\% / \textbf{95.00}\%             \\
CFTE~\cite{cfte} / Ours              &   4.29 / 4.21      &    0.20 / 0.14   &  0.31 / 0.23      &   850.03 / 390.17  &23.10 / 24.74 &   6.00\% / \textbf{94.00}\%             \\
LIA~\cite{lia} / Ours               &  4.34 / 4.21       &     0.15 / 0.14   &  0.26 / 0.23      &  467.19 / 390.17   &23.99 / 24.74 &  35.50\% / \textbf{64.50}\%              \\ \hline
\end{tabular}
\vspace{-0.2cm}
\end{table*}

Fig.~\ref{fig:11} provides visual quality comparisons of two particular sequences from moving-body and talking-face scenarios among all algorithms. It can be observed that our method can deliver the most visual-pleasing and temporal-coherent reconstructions for both scenarios. Specifically, at very low bit rates of 5kbps, the VVC reconstructed results exist severe blocking artifacts and their signal fidelity cannot be guaranteed for these two scenarios. As for other generative codecs, MRAA~\cite{mraa} and TPSM~\cite{tpsm} also suffer from blurring artifacts and poor texture details on moving-body and talking-face scenarios. In particular, the reconstruction quality of TPSM~\cite{tpsm} on the talking-face scenario cannot be perceived due to obvious visual artifacts. Besides, LIA~\cite{lia} is faced with obvious distortions with larger movements for both scenarios, and CFTE~\cite{cfte} cannot achieve accurate experession movements in mouth and eyes for talking-face scenario.



Furthermore, we conduct user study to compare our single-resolution model with all other algorithms at similar coding bit-rate. Specifically, we choose 20 sequences~(10 sequences from each test set), and implement ``two alternatives, force choice''~(2AFC) subjective test with 10 participants. During the test, these selected sequences from our method and other compared algorithms are sequentially displayed in a pair-wise manner, and the participants are asked to choose one video from each pair with better quality. To avoid experimental bias, we mix up all video pairs and randomly display them.

As shown in Table~\ref{tab:5}, these participants are more inclined to choose our reconstructed videos as the preferred video compared to other reconstruction results. In particular, our proposed method shows absolute advantage with more than 90\% preference ratio against VVC~\cite{vvc}, MRAA~\cite{mraa}, TPSM~\cite{tpsm} and CFTE~\cite{cfte}. As for the user preference between LIA and ours, our reconstructed videos can still be voted with a higher ratio 64.50\%. In addition, we also provide average results of these tested sequences in terms of perceptual-level measures DISTS, LPIPS, FVD as well as pixel-level measurement PSNR. At similar bit-rate, our method can achieve advantageous objective quality compared with other methods.

\begin{figure}[t]
    \centering
    \includegraphics[width=7cm]{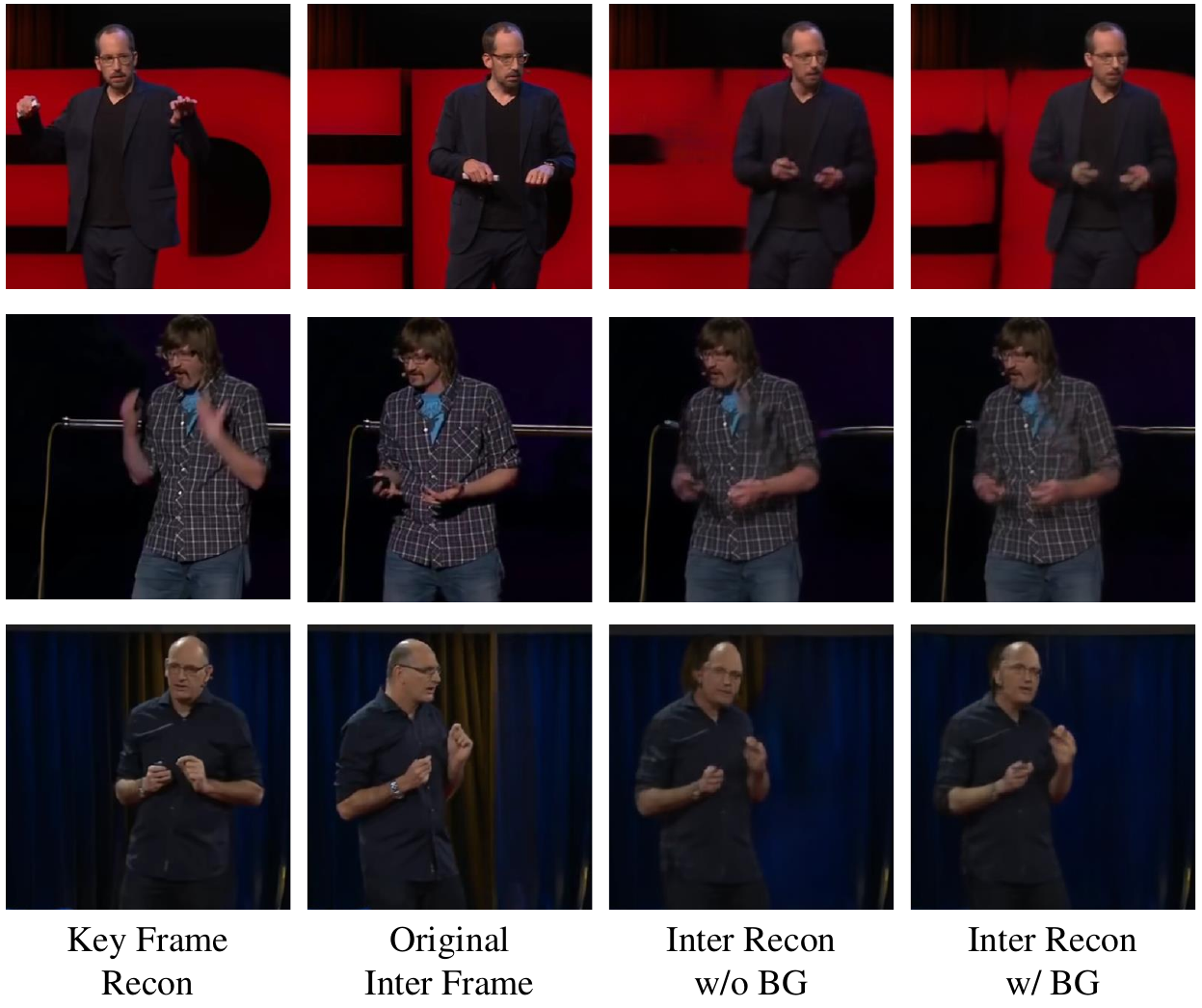}
    \caption{Subjective examples of independent background generation.}
    \vspace{-0.26cm}
    \label{fig:10}
\end{figure}


\begin{figure}[t]
\centering
\subfloat[Rate-DISTS]
{\includegraphics[width=4.5cm]{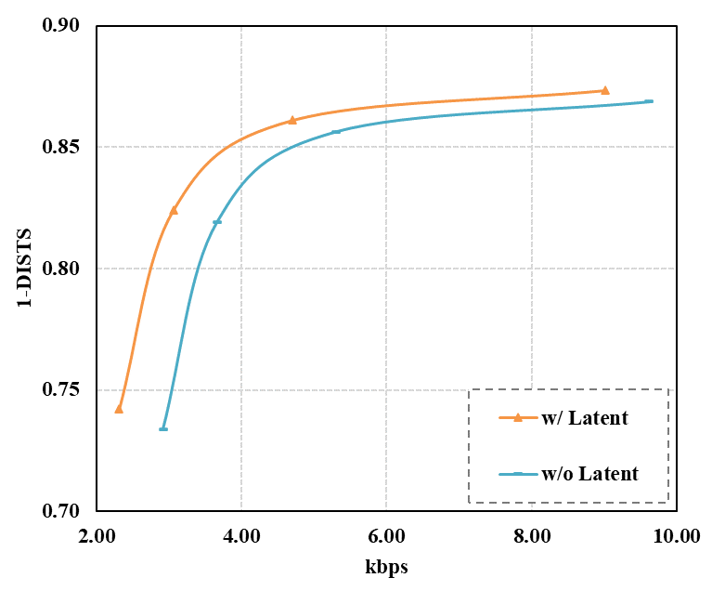}} 
\subfloat[Rate-LPIPS]{\includegraphics[width=4.5cm]{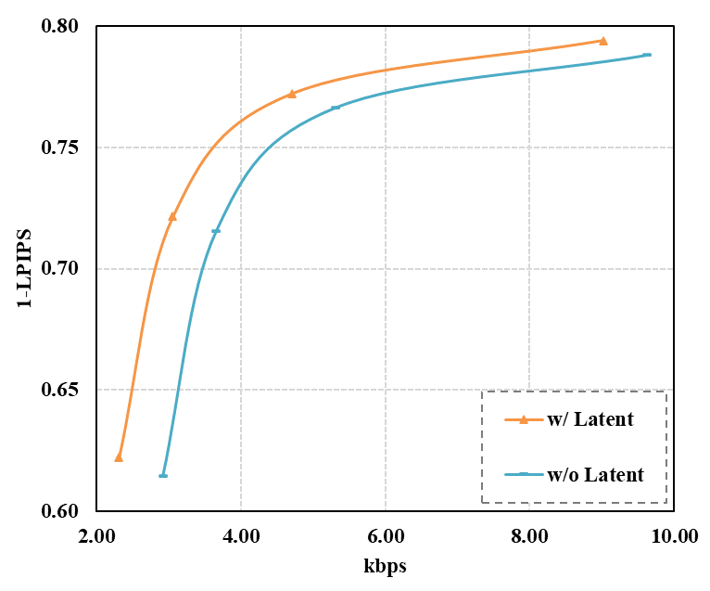}}

 \caption{RD performance comparisons for ablation study on latent feature. }
 \vspace{-0.26cm}
 \label{fig:8}
\end{figure}

\begin{figure}[t]
\centering
\subfloat[Rate-DISTS]
{\includegraphics[width=4.5cm]{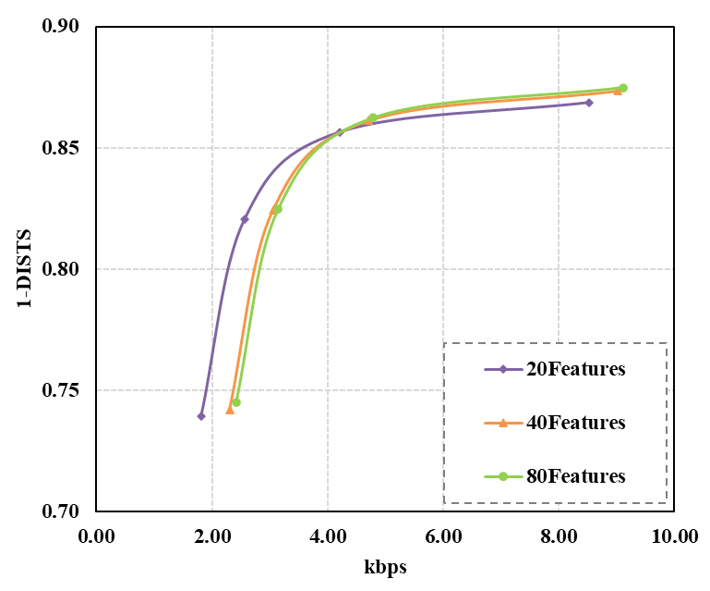}} 
\subfloat[Rate-LPIPS]{\includegraphics[width=4.5cm]{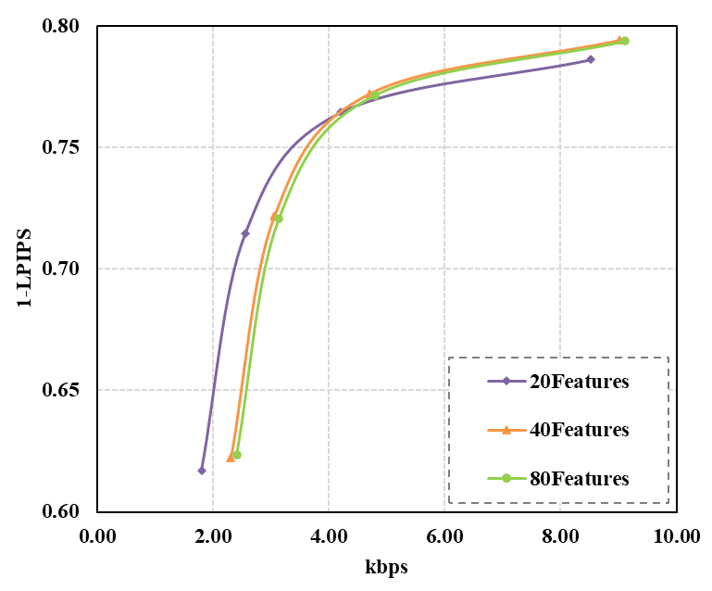}}

 \caption{RD performance comparisons for ablation study on the number of factorized features. }
 \vspace{-0.26cm}
 \label{fig:9}
\end{figure}

\subsubsection{Multi-resolution Coding Performance}
For multi-resolution coding, we use our proposed multi-resolution models to compare with VVC~\cite{vvc}. Specifically, we use Lanczos interpolation to resize our test set to different resolutions. For talking-face test set, 128$\times$128 and 512$\times$512 sequences are interpolated to suit our multi-resolution setting; For moving-body test set, 192$\times$192 and 768$\times$768 sequences are interpolated. Then, we use our multi-resolution model~(MTTF-multiRes) to evaluate all sequences. Table~\ref{tab:3} and Table~\ref{tab:4} show the multi-resolution compression performance in term of average BD-rate saving against VVC on moving-body and talking-face test set respectively. Experimental results demonstrate that our multi-resolution models can outperform VVC in a large margin and be able to achieve more than 60\% BD-rate saving in all resolutions with only one model for each scenario. In other words, the proposed resolution-expandable generator in the MTTF framework has high robustness and adaptivity, facilitating to achieve superior performance across different resolutions and video contents.

\subsection{Ablation Studies}
In this section, we execute the ablation studies on architecture designs and hyper-parameter settings in our framework. It should be noted that all following models are trained and evaluated on moving-body scenario setting.

\subsubsection{Independent Background Generation} In Section~\ref{Motion Estimation} and Section~\ref{generation}, we design a foreground-and-background parallel generation scheme that can employ these estimated motions to achieve independent content generation. Such scheme can bring more stable generation of occluded area and prevent incorrect attachment of foreground/background. To verify this, we use single-resolution model without background modelling~(denotes as ``w/o BG'') to compare with MTTF-singleRes model~(denotes as ``w/ BG''). As shown in Fig.~\ref{fig:10}, we provide subjective examples in terms of the reconstructed key frame, the original inter frame and one particular reconstructed inter frame with or without independent background generation. From the first two rows, the visual areas that is occluded by the body~(for example, letter ``E'' in the first row) in the key frame can be better predicted with independent generation. From the third row, the head of the body is incorrectly attached to background when the proposed method does not include independent background generation scheme. On the contrary, the proposed foreground-and-background parallel generation scheme can achieve better visual reconstruction and less occlusion artifact.

\subsubsection{Latent Feature in Multi-granularity Factorization}
In section~\ref{Multi-granularity Temporal Trajectory Factorization}, we introduce our proposed multi-granularity temporal trajectory factorization. One of the key insights is that this scheme utilizes spatial latent feature of reconstructed key frame as the basis of multi-granularity motion transformation to obtain the fine-grained motion fields. We claim that leveraging latent feature can provide more content information and improve the expressibility of input-adaptive motion fields. To verify this, we remove latent feature and compare the RD performance with corresponding anchor model. Specifically, we use single-resolution model without independent background generation as anchor model~(denotes as ``w/ Latent'') to avoid influences from resolution-expandable generator and independent generation. Then, we replace latent feature by a ``ones\_like'' tensor with the same dimension, so that only compact motion vector is obtained from key frame reconstructions~(denotes as w/o Latent). As shown in Fig.~\ref{fig:8}, the overall RD performance is dramatically degraded after removing latent features, which can indicate the effectiveness of the proposed feature factorization. 
\subsubsection{Number of Features in Factorized Motion}
In Section~\ref{Multi-granularity Temporal Trajectory Factorization} and Section~\ref{Motion Estimation}, we obtain $N_{F}$ features from multi-granularity factorization and estimate totally $2N_{F}$ motion components. When transmitted inter frame features, $2N_{F}$ parameters ($N_{F}$ weights and $N_{F}$ biases) are coded. Herein, we adjust the number of features $N_{F}$ and see their influences on RD performances. Similarly, we use single-resolution model without independent background generation as anchor model~(denotes as ``20 Features''). We then set number of features as 10 and 40 and compare their RD performances~(denotes as ``10 Features'' and ``40 Features'' respectively). As shown in Fig.~\ref{fig:9}, by using fewer features, the bit-rate decreases while quality drops. When increasing the number of features to 40, the RD performance barely changes, which means the features are saturated and further increasing its number would cause redundancy. To make a balance between bit-rate cost and reconstruction quality, we choose ``20 Features'' as our default setting in our experiments. 

\section{Conclusion}
In this paper, we propose a multi-granularity temporal trajectory factorization scheme for generative human video compression. By exploring the internal correlations between compact motion vectors and fine-grained motion fields, the proposed framework can well guarantee both signal compressiblity for ultra-low bit-rate and motion expressibility for high-fidelity reconstruction. Furthermore, a resolution-expandable and foreground-background-parallel generator is designed to improve the generalizability and flexibility of the proposed generative codec across different human video contents and different resolutions. Experimental results show that our proposed framework can outperform both state-of-the-art conventional video codec with more than 70\% BD-rate saving, as well as existing generative codecs with large margins. 


\bibliographystyle{IEEEtran}
\bibliography{references}
\vspace{-1cm}
\begin{IEEEbiography}[{\includegraphics[width=1in,height=1.25in,clip,keepaspectratio]{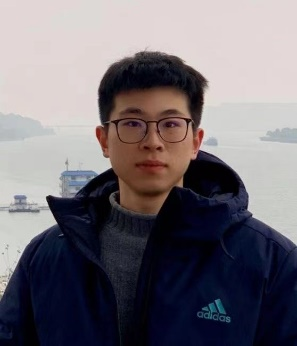}}]{Shanzhi Yin}  received the B.E. degree in communication engineering from Wuhan University of
Technology, Wuhan, China, in 2020, and the M.S.
degree in information and communication engineering from Harbin Institute of Technology, Shenzhen,
China, in 2023. He is currently pursuing the Ph.D.
degree with the Department of Computer Science,
City University of Hong Kong. His research interests
include video compression and generation.
\end{IEEEbiography}
\vspace{-1cm}
\begin{IEEEbiography}[{\includegraphics[width=1in,height=1.25in,clip,keepaspectratio]{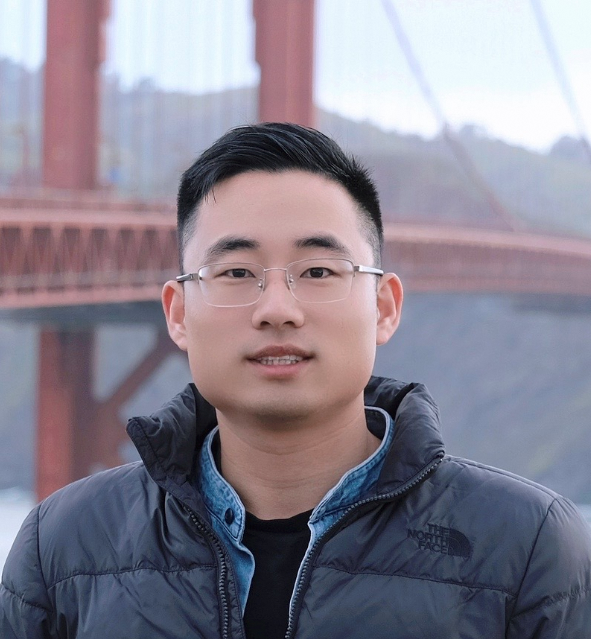}}]{Bolin Chen} received the B.S. degree in communication engineering from Fuzhou University, Fuzhou, China, in 2020. He is currently pursuing the Ph.D. degree with the Department of Computer Science, City University of Hong Kong. His research interests include video compression, generation and quality assessment. 
\end{IEEEbiography}
\vspace{-1cm}
\begin{IEEEbiography}[{\includegraphics[width=1in,height=1.25in,clip,keepaspectratio]{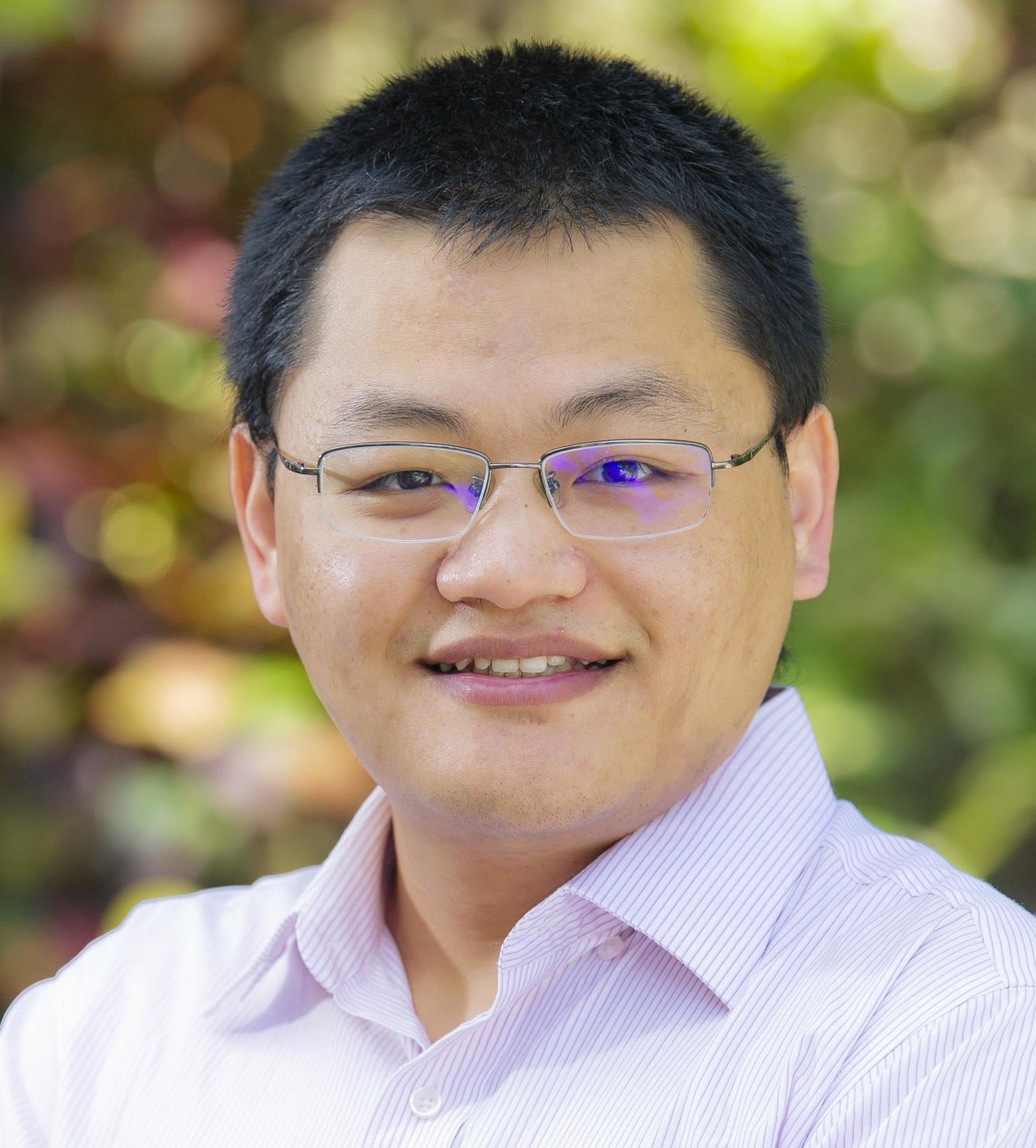}}]{Shiqi Wang} (Senior Member, IEEE) received the B.S. degree in computer science from the Harbin Institute of Technology in 2008 and the Ph.D. degree in computer application technology from Peking University in 2014. From 2014 to 2016, he was a Post-Doctoral Fellow with the Department of Electrical and Computer Engineering, University of Waterloo, Waterloo, ON, Canada. From 2016 to 2017, he was a Research Fellow with the Rapid-Rich Object Search Laboratory, Nanyang Technological University, Singapore. He is currently an Assistant Professor with the Department of Computer Science, City University of Hong Kong. He has proposed more than 50 technical proposals to ISO/MPEG, ITU-T, and AVS standards, and authored or coauthored more than 200 refereed journal articles/conference papers. His research interests include video compression, image/video quality assessment, and image/video search and analysis. He received the Best Paper Award from IEEE VCIP 2019, ICME 2019, IEEE Multimedia 2018, and PCM 2017. His coauthored article received the Best Student Paper Award in the IEEE ICIP 2018. He was a recipient of the 2021 IEEE Multimedia Rising Star Award in ICME 2021. He serves as an Associate Editor for \textsc{IEEE Transactions on Circuits and Systems for Video Technology}. 
\end{IEEEbiography}
\vspace{-1cm}
\begin{IEEEbiography}[{\includegraphics[width=1in,height=1.25in,clip,keepaspectratio]{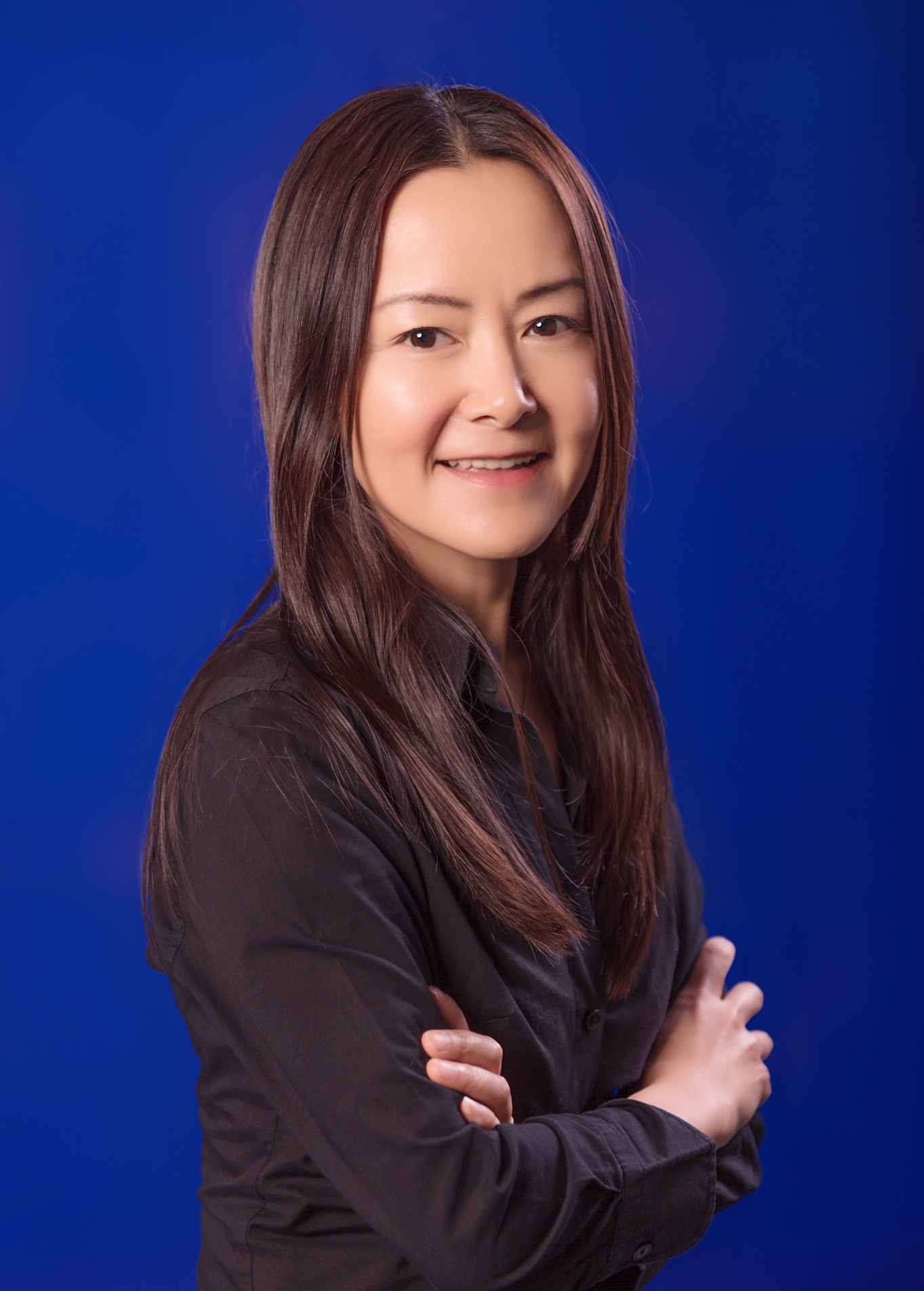}}]{Yan Ye} (Senior Member, IEEE) received the B.S. and M.S. degrees in electrical engineering from the University of Science and Technology of China in 1994 and 1997, respectively, and the Ph.D. degree in electrical engineering from the University of California at San Diego, in 2002. She is currently the head of Video Technology Lab at Alibaba Damo Academy, Alibaba Group U.S., Sunnyvale, CA, USA, where she oversees multimedia standards development, video codec implementation, and AI-based video research. Prior to Alibaba, she was with the Research and Development Labs, InterDigital Communications, Image Technology Research, Dolby Laboratories, and Multimedia Research and Development and Standards, Qualcomm Technologies, Inc. She has been involved in the development of various video coding and streaming standards, including H.266/VVC, H.265/HEVC, scalable extension of H.264/MPEG-4 AVC, MPEG DASH, and MPEG OMAF. She has published more than 60 papers in peer-reviewed journals and conferences. Her research interests include advanced video coding, processing and streaming algorithms, real-time and immersive video communications, AR/VR, and deep learning-based video coding, processing, and quality assessment. 
\end{IEEEbiography}

\end{document}